\documentclass[twocolumn]{aastex63}
\usepackage{amssymb,bm,graphicx,verbatim,soul}
\usepackage{natbib,tikz,lipsum}
\tikzset{>=latex}

\setcitestyle{notesep={ }} % tells natbib not to insert comma before
\newcommand{\arcs}{\ensuremath{^{\prime\prime}}}

\newcommand{\dproj}{\ensuremath{d_{\rm proj}}}
\newcommand{\dv}{\ensuremath{\Delta v_{\rm los}}}
\newcommand{\dlol}{\frac{\Delta \lambda}{\lambda}}

\begin{document}
\title{Toward Detecting the Moving Lens Effect with Optical Spectroscopy}
\shorttitle{Moving Lens Effect}

\author[0000-0002-0813-5888]{David Wittman}
\affiliation{Department of Physics and Astronomy, University of California, Davis, CA 
  95616 USA}
% Rodrigo designed the masks and led the observations
\author[0000-0002-6217-4861]{Rodrigo Stancioli}
\affiliation{Department of Physics and Astronomy, University of California, Davis, CA 95616 USA}
\author[0000-0001-8559-4538]{Christopher Hopp}
\affiliation{Department of Physics and Astronomy, University of California, Davis, CA 
  95616 USA}


% Re-edit abstract last!
\begin{abstract} Relativity predicts that sky-plane motion of a
  gravitational lens changes the wavelengths of lensed photons.  This
  moving lens effect is orders of magnitude smaller than cosmological
  and Doppler effects on the same photons, but can be isolated by
  comparing different strongly lensed images of a given source.  With
  a massive, fast-moving lens such as a galaxy cluster falling into
  another cluster, lensed spectra of a given source could differ in
  wavelength by up to 1 part in $10^6$.  We examine the feasibility of
  detecting this effect with optical spectroscopy, using Keck/Deimos
  data on the merging cluster RM J001938.7+033557.3 as a case study.
  We also quantify the predicted sky-plane motions of systems likely
  to be identified as merging clusters in a $\Lambda$CDM cosmology,
  and investigate how information on sky-plane motions could help
  constrain merger models.  We find that measurement of the moving
  lens effect would disambiguate between models in which the
  subclusters are outbound after first pericenter and those in which
  the subclusters have returned to the same separation after first
  apocenter.
\end{abstract}

\section{Introduction}\label{sec-intro}

%  inverted triangle para
Galaxy velocities are foundational to astrophysics, yet astronomers
can measure only one of three velocity components---the radial
velocity via the Doppler effect.  There are exceptions: the proper
motions of Local Group galaxies can now be measured by comparing sky
positions measured many years apart \citep[][and references
therein]{Bennet-PM-2024}.  But beyond the local universe, astronomers
must make do with measuring one velocity component, which limits the
ability to model orbits.

A specific case of orbit modeling for extremely distant objects is the
study of merging galaxy clusters.  With masses of ${\sim}10^{15}$
M$_\odot$ falling from separations of ${\sim}10$ Mpc, they reach
pericenter speeds of ${\sim}3000$ km/s.  They then take ${\sim}1$ Gyr
to oscillate through a first apocenter (at a separation of ${\sim}1$
Mpc) to a second pericenter, followed by further damped oscillations;
see, e.g., \citet{Kim2017} for clear illustrations spanning various
initial speeds and dark matter models. Modeling the orbit, at least
through the first oscillation, is a key ingredient for studying
possible momentum transfer between dark matter particles
\citep{Markevitch04,Randall2008,Robertson2017Bullet}, merger effects on
intracluster medium (ICM) properties
\citep{ZuHone2011,Skillman13,ZuHone2018}, and the generation and
propagation of shocks therein
\citep{Reinout2011CIZA,BotteonA115,ReinoutRadioReview19}.

The basic observables available to constrain such an orbit are the
projected separation between the subclusters, \dproj; their
line-of-sight (LOS) velocity difference \dv; and the masses $M_1$ and
$M_2$.  However, because we are unable to distinguish which subcluster
is closer to us, the sign of \dv\ carries no information regarding
whether the subclusters are approaching or receding from each other.
This leads to a discrete degeneracy: with a fixed set of observables
$(\dproj,\dv,M_1,M_2)$, the subclusters could be inbound toward
pericenter, outbound after pericenter, or returning after apocenter.
To some extent this degeneracy can be broken by other observables such
as X-ray properties and shock position.  The X-ray temperature jumps
dramatically around first pericenter and then cools
\citep{ZuHone2011,ZuHone2018}, but lack of knowledge of gas conditions
prior to the merger creates substantial uncertainty in this
inference. The shock is launched around pericenter and moves outward
monotonically, thus providing a sort of clock
\citep{Randall2008,Wonki2026relics}, but uncertainties in propagation
speed and projection effects can render this indecisive
\citep[e.g.,][]{Ng2015}.  

In this context, a plane-of-sky (POS) velocity measurement would be a
welcome addition to the list of observables.  The moving lens effect
\citep{Birkinshaw1983} has been proposed for this purpose
\citep{MolnarBirkinshaw2003}.  \citet{Birkinshaw1983} considered a
scenario where source and observer share a frame through which a lens
moves with sky-plane velocity $\bm{v}$. To restate their Equation~9 in
modern lensing notation, they showed that if a lens provides a
deflection $\hat{\bm{\alpha}}$ in its frame, Lorentz transformations
result in the photon losing a fractional energy
$\gamma\frac{\bm{v}}{c}\cdot\hat{\bm{\alpha}}$ (where $c$ is the speed
of light, $\gamma$ is the Lorentz factor, and bolded symbols represent
2-D sky-plane vectors) in the observer frame.  Known gravitational
lenses travel at $v\ll c$ for which $\gamma$ departs negligibly from
one, hence
%\begin{equation}
%  \label{eq:1}
%  \frac{\Delta E}{E}  =  -\frac{\bm{v}}{c}\cdot\hat{\bm{\alpha}}
%\end{equation}
%or
\begin{equation}
  \label{eq:2}
  \frac{\lambda_{\rm obs}}{\lambda_{\rm s}} = 1+ \frac{\bm{v}}{c}\cdot\hat{\bm{\alpha}}
\end{equation}
is a good approximation for the observed wavelength
given $\lambda_s$, the wavelength that would have been
observed for a static lens.  \citet{Molnar2013} provide a more thorough
summary of the literature on this effect. They conclude that tangential
motions of observer and source can be neglected in this context,
largely because merging clusters achieve much higher peculiar
velocities than plausibly expected for source or observer, and
partially because in the full expression derived by \citet{Wucknitz2004}
the latter motions are multiplied by distance ratio factors
less than unity.

The effect is small, but has recently been detected in a different
context, via the resulting dipole pattern aligned along the direction
of the lens velocity in cosmic microwave background
(CMB) temperature maps \citep{2026arXiv260518938H}. For galaxy
clusters, with $\sim30\arcs$ deflections, the angular size of this
dipole is potentially confused with primary CMB anisotropies. Galaxies
provide smaller deflections ($\sim1\arcs$), reducing this
confusion. Each galaxy provides a correspondingly small
$\frac{\Delta T}{T}$, but \citet{2026arXiv260518938H} used a very
large galaxy sample to recover a statistical detection.  This work
explores the very different context of single-object detections.  The
best lenses for such purposes will be those with large deflection
angles as well as high-speed.  This is a good match to the merging
cluster scenario, where speeds can reach 3000 km/s for a $10^{-6}$
fractional effect.  Furthermore, we consider the use of background
galaxies rather than the CMB as sources: although they appear along
far fewer lines of sight, they provide well-defined spectral lines for
precise spectroscopy.

\begin{figure}
\centerline{
\begin{tikzpicture}
  \node[circle,fill=lightgray] at (-1.5,0) {$M_1$};
  \draw[->] (-1.9,0) -- +(-0.5,0);
  \node[circle,fill=lightgray] at (1.5,0) {$M_2$};
  \draw[->] (1.9,0) -- +(0.5,0);
  \node at (-2.5,0) {\textcolor{red}{A}};
  \node at (-0.5,0) {\textcolor{blue}{B}};
  \node at (-1.5,1) {C};
  \node at (-1.5,-1) {D};
  \node at (2.5,0) {\textcolor{red}{E}};
\end{tikzpicture}}
\caption{Hypothetical lensed image configurations and lens motions for
  a head-on cluster merger seen after first pericenter, and with a
  line of sight perpendicular to the orbit.  The
  wavelength shift of Image A relative to B would
be ideal for probing the velocity of $M_1$. Images at C and D would be
insensitive to the motion of $M_1$. If images of the same source
appeared at A and E, both images would be affected by the same sign
wavelength shift, so their spectra would be insensitive to the
relative motion of $M_1$ and $M_2$.}
\label{fig-cartoon} 
\end{figure}
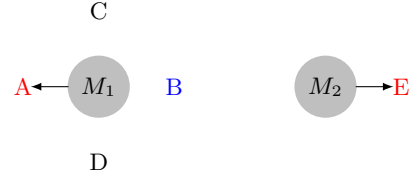

\citet{Molnar2013} considered multiple images of a strongly lensed
background galaxy in the context of the Bullet cluster. This naturally
provides a reference wavelength: absent the moving lens effect, all
images would be measured as having identical redshifts.
Consider images A and B in Figure~\ref{fig-cartoon}. Taking the ratio
of observed wavelengths eliminates $\lambda_s$, yielding the expression
\begin{eqnarray}
  \frac{\lambda_A}{\lambda_B}&=&\frac{1+\frac{\bm{v}}{c}\cdot\hat{\bm{\alpha}}_{\mathrm A}}{1+\frac{\bm{v}}{c}\cdot\hat{\bm{\alpha}}_{\mathrm B}}\\
                        &\approx&
                                  1+\frac{\bm{v}}{c}\cdot(\hat{\bm{\alpha}}_{\mathrm A}-\hat{\bm{\alpha}}_{\mathrm B})   \label{eq:3}
\end{eqnarray}
which is independent of source properties including cosmological
redshift.  We quantify the effect in terms of
$\frac{\lambda_{\mathrm A}}{\lambda_{\mathrm B}}-1\equiv \dlol$, which
is zero for a static lens.\footnote{To avoid confusion with other
  physical effects, we discourage parametrization of this wavelength
  difference as a velocity or redshift.} In the standard lensing
formalism, $\bm{\hat{\alpha}}$ enters observations only via the
``reduced'' deflection angle
$\bm{\alpha}\equiv \bm{\theta}-\bm{\beta}$, where $\bm{\theta}$ is the
observed sky position and $\bm{\beta}$ is the (unobservable) sky
position in the absence of lensing. The two deflection angles are
related via
$\bm{\alpha}=\frac{D_{\mathrm ds}}{D_{\mathrm s}}\hat{\bm{\alpha}}$
where $D_{\mathrm ds}$ and $D_{\mathrm s}$ are the angular diameter
distances from deflector to source and from observer to source
respectively.  Two images from the same source must have the same
$\bm{\beta}$, so the difference in observed positions
\begin{eqnarray}
  \Delta\bm{\theta} &\equiv &\bm{\theta}_{\mathrm A}-\bm{\theta}_{\mathrm B}\\
                    &=&\bm{\alpha}_{\mathrm A}-\bm{\alpha}_{\mathrm B}\\
  &=&\frac{D_{\mathrm ds}}{D_{\mathrm s}}(\hat{\bm{\alpha}}_{\mathrm A}-\hat{\bm{\alpha}}_{\mathrm B}).   \label{eq:7}
\end{eqnarray}
Combining Equations~\ref{eq:3} and ~\ref{eq:7} yields
\begin{equation}\label{eq-dlol}
  \dlol = \frac{D_{\mathrm s}}{D_{\mathrm ds}}\frac{\bm{v}}{c}\cdot\Delta\bm{\theta}.
\end{equation}
Hence $\bm{v}\cdot\Delta\bm{\theta}$ may be inferred without reference
to a lens model, although in \S\ref{ssec-confound} we consider how
observational difficulties may require a lens model in practice.

Returning to Figure~\ref{fig-cartoon}, we now consider the other image
pairs. The lens motion shown does not affect photon energies in images
C and D. The motion of $M_2$ affects photon energies in image E the
same way that the motion of $M_1$ affects image A, so an A-E
comparison is not probative.  In summary, image pairs split along the
direction of motion of a single subcluster, such as A-B, provide the
largest $\dlol$ (at fixed image separation).  However, these are rare:
the two overlapping, extended masses in a binary cluster merger
approximate a bar of mass, which tends to split images in the
direction perpendicular to the bar. So observed image pairs in merging
clusters are most often like the C-D configuration. However, these
pairs may be more useful than suggested by this toy model which
\textit{assumes} a radial trajectory. The C-D configuration can be
used to empirically probe the azimuthal component of the trajectory
and constrain the orbit model. We show in this paper that POS speeds
in the direction transverse to the apparent separation vector can
reach many hundreds of km/s, yielding C-D $\dlol$ signals smaller than
the $10^{-6}$ quoted above, but only by factors of several. This is
more challenging to detect, but the plethora of transverse image pairs
may offer opportunities to make useful constraints on the transverse
motion, if not detections.

We have discovered a quadruply imaged background galaxy behind one
subcluster of the merging cluster RM J001938.7+033557.3.  The images
are bright enough to be observed with ground-based optical
spectroscopy at spectral resolution $R\approx6000$.  We present
spectra of two of the images and analyze them to study the feasibility
of detecting the moving lens effect. The remainder of the paper is
organized as follows: \S\ref{sec-data} describes the data,
\S\ref{sec-analysis} analyzes the spectra, \S\ref{sec-mergermodeling}
quantifies the cosmologically expected range of POS velocities (both
parallel and perpendicular to the projected subcluster separation
vector), and \S\ref{sec-discuss} summarizes with a discussion of the
limitations of the current work and directions for future work.

\section{Data}\label{sec-data}

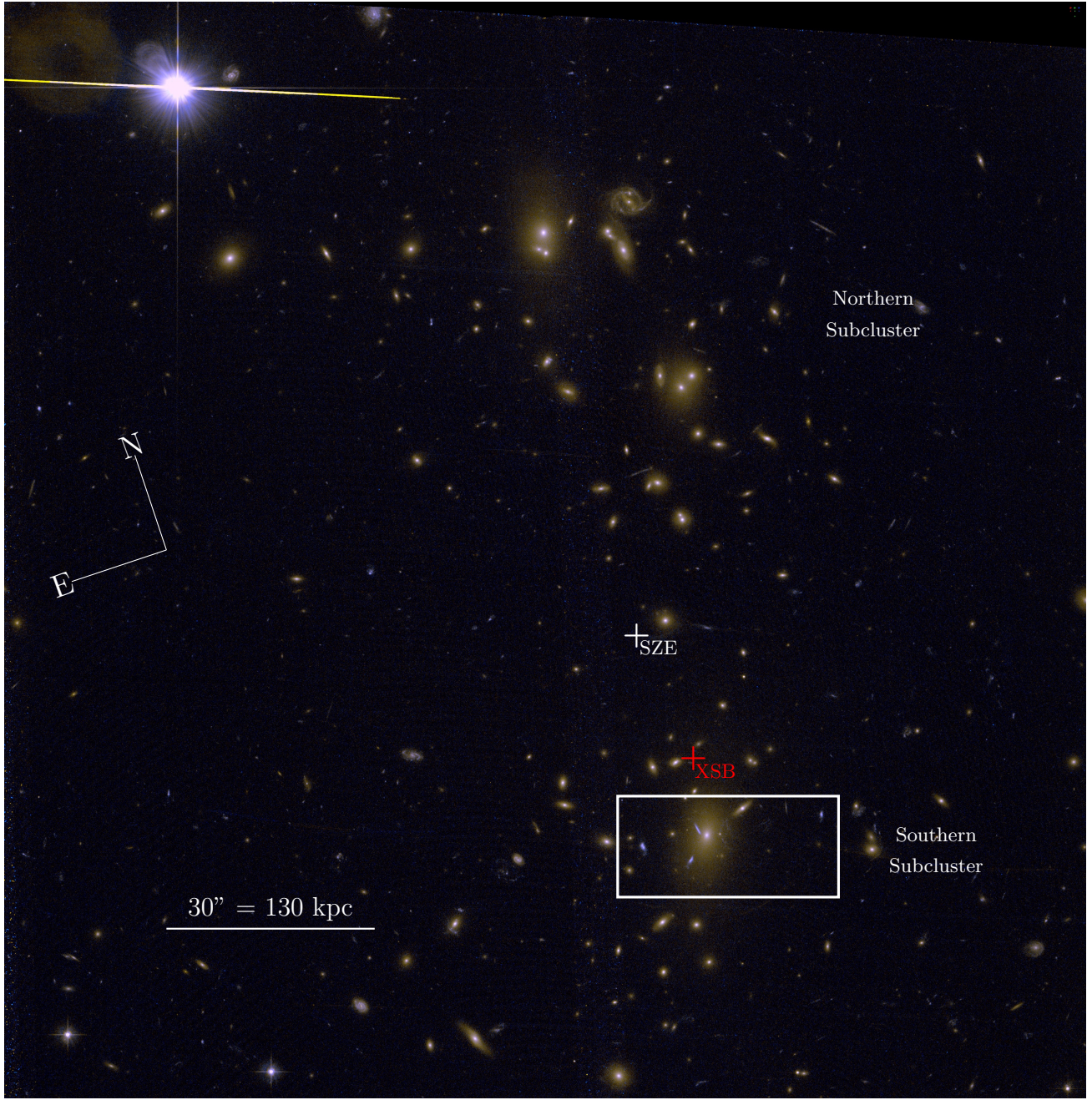
\begin{figure*}
\centerline{
\begin{tikzpicture}
\node at (0,0) {\includegraphics[width=0.95\textwidth]{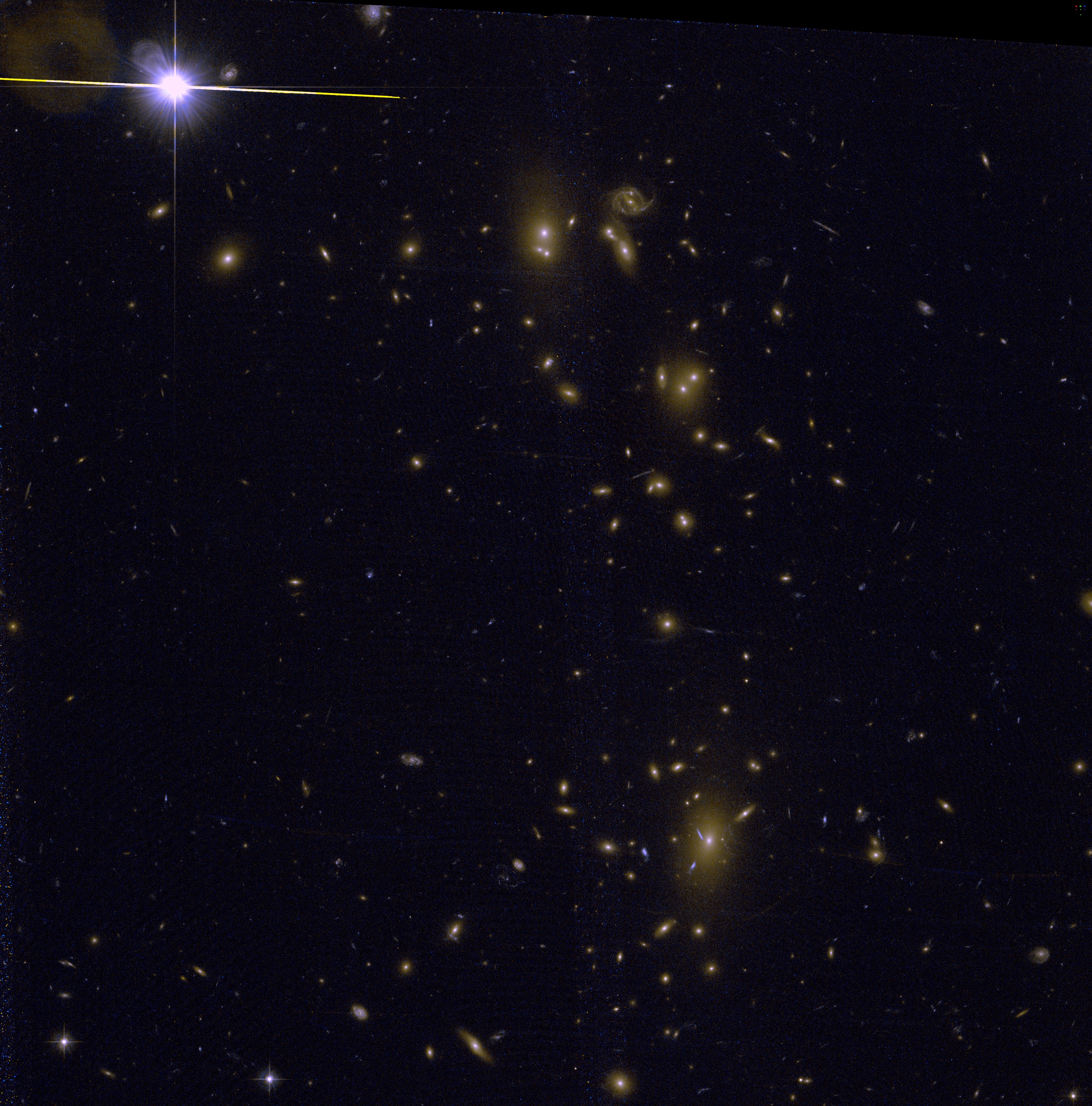}};
\draw[white,very thick] (1.15,-3.9) rectangle +(3.5,-1.6);
\node at (2.35,-3.3) {\Large\textcolor{red}{\bf +}};
\node at (2.7,-3.5) {\textcolor{red}{XSB}};
\node at (1.45,-1.35) {\Large\textcolor{white}{\bf +}};
\node at (1.8,-1.55) {\textcolor{white}{SZE}};
\draw[white] (-6,0) to node[pos=1.1,rotate=20] {\textcolor{white}{\Large N}}  +(-0.5,1.5);
\draw[white] (-6,0) to node[pos=1.1,rotate=20] {\textcolor{white}{\Large E}} +(-1.5,-0.5);
\draw[white,thick] (-6,-6) to node[above,pos=0.5] {\textcolor{white}{\large 30'' = 130 kpc}} +(3.3,0);
\node at (6.2,-4.5) {\textcolor{white}{Southern}};
\node at (6.2,-5) {\textcolor{white}{Subcluster}};
\node at (5.2,4) {\textcolor{white}{Northern}};
\node at (5.2,3.5) {\textcolor{white}{Subcluster}};
\end{tikzpicture}}
\caption{HST color image showing the two subclusters of RM J0019. The
  white rectangle shows the area covered by
  Figure~\ref{fig-fourimages}. The red cross marks the
  XSB peak, and the white cross marks the main SZE peak in the ACT
  DR6 catalog \citep{ACT-DR6-clustercat}. That work also found a
  secondary peak off the south edge of this image.}
\label{fig-bigview}
\end{figure*}

We identified RM J001938.7+033557.3 ($z=0.27$) as a potential merging
cluster on the basis of its optical binarity, as quantified by the
separation of its top brightest cluster galaxy (BCG) candidates in the
redMaPPer \citep{Rykoff2014,Rykoff2016} catalog based on SDSS
photometry.  The optical selection method is described in more detail
in \citet{HoppXSORTER}. With a redMaPPer richness of 109, this cluster
failed to make the cut for further analysis in that paper, but it is
listed in Table 4 of that paper as a less-rich cluster worthy of
followup.  It appears in the XMM-\textit{Newton} archive with 48 ks of
exposure time (Observation ID 0693010301 in 2012) as part of a project
to examine scaling relations in a large sample of clusters (PI:
Y.-Y. Zhang).  We found that the X-ray surface brightness (XSB) peak
is located between the two galaxy subclusters, suggesting that the two
subclusters have had a recent first pericenter passage. We then
obtained HST/ACS data (Proposal ID 17114, PI: D. Wittman), consisting
of 2193 seconds of exposure time in F606W and 2208 in F475W. The
resulting color image (where blue/red is F475W/F606W and green is the
sum of the two) is shown in Figure~\ref{fig-bigview}.  See Stancioli
et al (in prep) for a more detailed analysis of the merging scenario
using the XMM-\textit{Newton} and HST/ACS data as well as a
Keck/Deimos galaxy redshift survey. For this work, it is sufficient to
note that the two subclusters are physically related rather than a
chance projection, and their small projected separation suggests a
phase near pericenter where the relative speed is high.

Independently, \cite{Meerkat2021analysis} also identified these
subclusters as forming a single system. They conclude that it is
observed after a pericenter passage based on three lines of evidence:
an offset between the XSB peak and centroid (using the same
XMM-\textit{Newton} archival data used to trigger our HST followup);
an offset between the XSB peak and the Sunyaev-Zeldovich effect (SZE)
peak as measured by \citet{Hilton_2018}\footnote{The SZE detection was
  made using the Atacama Cosmology Telescope (ACT), hence these papers
  refer to the cluster as ACT-CL J0019.6+0336.}; and diffuse radio
emission detected by MeerKAT \citep{Meerkat2021}.

\begin{figure*}
\centerline{\includegraphics[width=\textwidth]{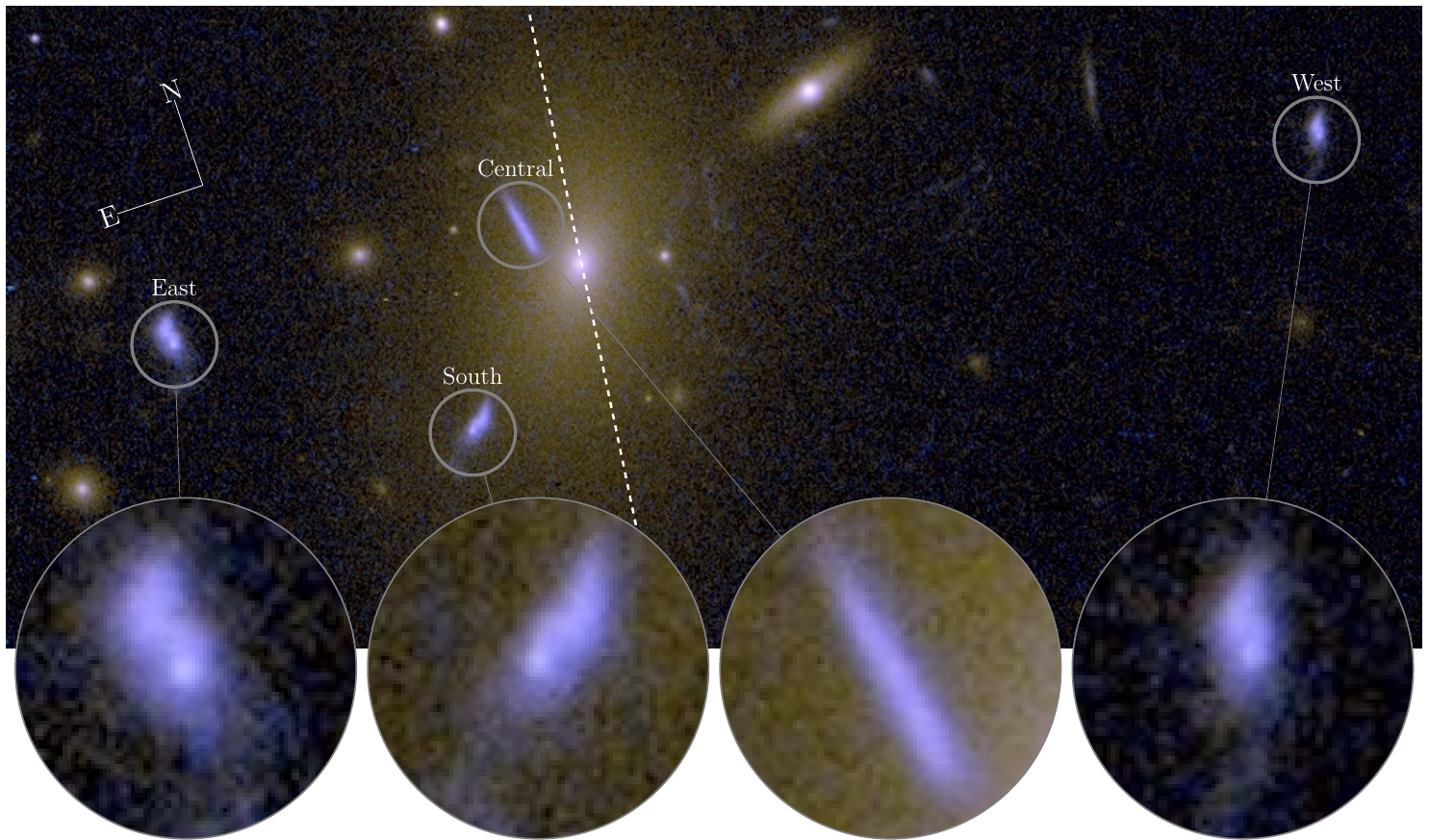}}
\caption{Magnified view of the white box in Figure~\ref{fig-bigview}.
  The rectangle is 40'' (170 kpc) wide. The dashed line shows the
  direction of a hypothetical radial orbit; the northern subcluster is
  well off the top of the image.}
\label{fig-fourimages}
\end{figure*}

We now show that this system has a bright multiply-imaged background
galaxy, making it a good candidate for practical attempts to detect
the moving lens effect. Figure~\ref{fig-fourimages} shows the area
around the southern BCG, with insets highlighting the multiple
images. The east, south, and west images exhibit remarkably similar
morphology (roughly that of a question mark, with inverted parity for
the southern image), strongly suggesting that they are multiple lensed
images of a single source.  The morphology of the central image is
highly distorted compared to the other images, such that its status as
a fourth image is not immediately obvious. However, its matching color
and surface brightness make it a candidate for a fourth image, and the
spectroscopy described below confirms it.  The east, south, and west
images are bright enough to be detected even in SDSS imaging, with $r$
magnitudes ranging from 21.96 to 22.08. The HST/ACS imaging shows that
the central image is about equally bright, despite blending with the
BCG in ground-based surveys. The bright images afford an opportunity
to obtain unusually high signal-to-noise ratio (S/N) spectra of lensed
images.

% stopped editing here 7/16/26. I should rephrase given the idea of
% probing transverse motion
The ideal moving-lens target would feature a lensed source capable of
supporting a high-precision wavelength comparison, as well as a large
predicted $\dlol$. RM J001938.7+033557.3 scores highly in the first
category: the bright images enable high S/N spectra, and we show in
this work that the source has strong spectral lines. Its promise in
the second category is less certain. A radial orbit would follow the
dashed line in Figure~\ref{fig-fourimages} in either direction, such
that the central and south images straddle the BCG roughly along the
direction of motion.  The small separation between these images
(${\approx}6\arcs$) would then yield $\dlol$ only $4\times 10^{-7}$ at
the maximum speed expected from a major cluster merger (3000 km/s).
At the other extreme, orbital motion in the east-west direction
coupled with the much larger (${\approx}33\arcs$) east-west separation
would create a larger effect, $\dlol=2\times10^{-6}$ at that speed.
These are best-case scenarios because a random merger is unlikely to
be observed at a moment of maximum speed and with all its velocity in
the POS.

We obtained spectra of the east and central images with the Deimos
multi-object spectrograph \citep{FaberDEIMOS} at the W. M. Keck
Observatory on 14 September 2023 (UT). These were spare slits in our
investigation of member galaxy velocities (Stancioli \textit{et
  al.} in prep).  When the observations were designed, the nature of
the central image was in doubt because its shape is so distorted
compared to the distinctive morphologies of the other images.
Confirming the central image required obtaining its redshift as well
as that of one other image.  The east (central) image was assigned a
spare slit on mask \texttt{0019A} (\texttt{0019B}). Observations used
the 1200 line mm$^{-1}$ grating, which yields a pixel scale of 0.33
\AA\ pixel$^{-1}$ and a resolution of ${\sim}1$ \AA\ with 1\arcs\ wide
slits. The grating was tilted for a central wavelength of 5550 \AA,
thus covering the range 4200--6900 \AA\ for a typical slit.  The
exposure time for each mask was $3\times900$ seconds. The seeing
varied only slightly, from 0.74--0.86\arcsec, during the
observations. We calibrated and reduced the data to a series of 1-D
spectra using PypeIt \citep{pypeit:joss_pub,pypeit:zenodo}.  We
double-checked the arc lamp wavelength calibration against sky
emission lines, and found good agreement.  We followed the
standard procedure of coadding the three exposures of each mask (but
also explore modeling the raw data in \S\ref{sec-analysis}).

\begin{figure*}
\centerline{\includegraphics[width=\textwidth]{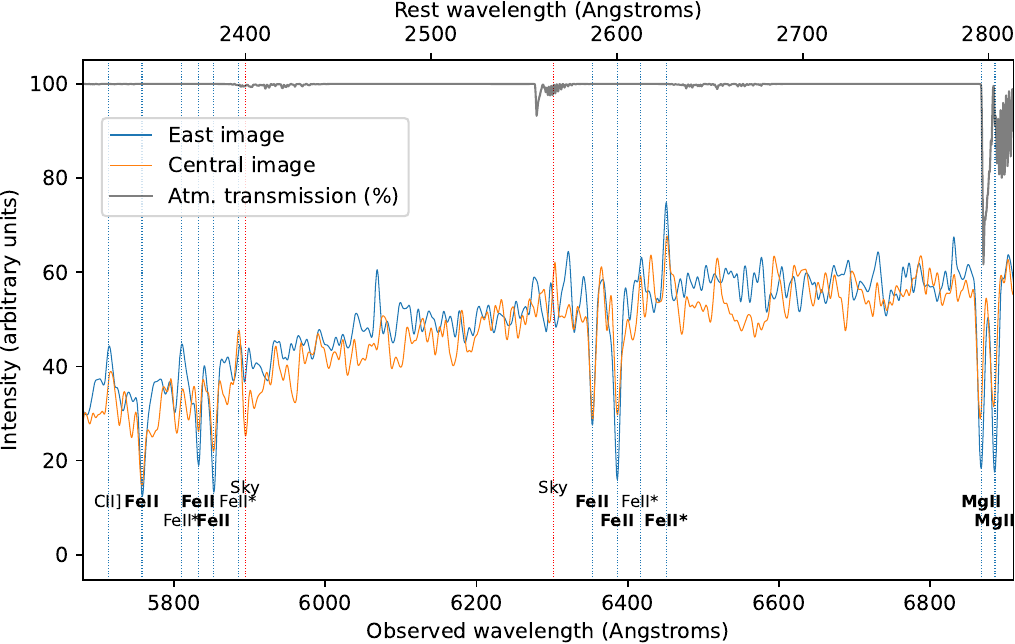}}
\caption{Spectra of the eastern and central images, focused on the
  region with clear spectral lines. The spectra are not flux
  calibrated so the overall shape of the continuum may reflect
  instrument throughput which varies slowly with wavelength.
  Atmospheric transmission (gray) and sky emission lines (red marks
  the brightest few) have been modeled out.  The rest frame wavelength
  axis reflects a redshift of 1.4561. The data have been smoothed with
  a Gaussian kernel of 8 pixel (2.6 \AA) width for illustration
  purposes. Lines with bold labels are those that are fit in
  \S\ref{sec-analysis}.}
\label{fig-spectra}
\end{figure*}

Figure~\ref{fig-spectra} shows the coadded spectra extracted from the
observations, focusing on the 5700--6900 \AA\ region where there are
clearly identifiable spectral features. The spectra are nearly
identical. The principal difference is that the absorption lines in
the spectrum of the central image are not quite as deep as those of
the east image.  This is likely due to incompletely subtracted
emission from the BCG; at ground-based resolution, the BCG overlaps
the central image.  A second difference is that the central image has
a continuum slightly fainter than that of the east image; this could
be due to higher lensing magnification of East and/or BCG absorption
of light from Central.

% margina) CII] $\lambda2326.00$ emission not worth mentioning
The spectra show seven identifiable absorption lines: FeII
$\lambda\lambda2344.21,2374.46,2382.76, 2586.65,2600.17$, and MgII
$\lambda\lambda2796.35,2803.53$, as well as FeII* $\lambda2626.45$
emission.  These will be described in further detail in
\S\ref{sec-analysis}.  Of relevance to this section, the lines appear
at a redshift of 1.4561 in each image. This demonstrates that the
central and east images are indeed of the same source galaxy.  A
working lens model suggests that the central image is actually
composed of two merging images, thus explaining the morphological
distortion (S. M. R. Adnan et al, in preparation). We proceed with an
analysis aimed at constraining the $\dlol$ factor relating the two
images.

\section{Analysis of lensed spectra}\label{sec-analysis}

In this section we quantify the width and signal-to-noise ratio (S/N)
of the observed spectral lines, and perform fits to both raw and
reduced data to highlight requirements for future observations that
could lead to detection of the moving lens effect.

% point of this subsection: initial assessment of info content based
% on std reduction, ignoring systematic. **Just make sure to frame it
% that way.**
\subsection{Statistical precision}\label{subsec-anal-stat}

% Present the line table w/widths, lead to basic stat prec. conclusions.
\textit{Basic line fitting with standard data products.} We fit a
Gaussian to each of the eight iron and magnesium lines listed near the
end of \S\ref{sec-data}, using the coadded spectra (as shown in
Figure~\ref{fig-spectra}, but without smoothing applied). Due to
overlap of the MgII lines, we fit those two lines
simultaneously. Table~\ref{tab:linefits-east} lists the results: the
typical statistical uncertainty on each line's central wavelength is
about 0.25 \AA, for a fractional wavelength uncertainty of about
$4\times10^{-5}$ per line.  As a best-case scenario a set of $n$ lines
can provide an uncertainty scaling as $n^{-1/2}$, but with eight lines
this yields only $1.4\times10^{-5}$ fractional wavelength uncertainty,
still nearly two orders of magnitude too large to detect the moving
lens effect. (The same fitting procedure applied to the central image
gives slightly lower S/N and slightly higher uncertainties due to BCG
light, but this affects only one of the four images in this system.)
Reaching fractional uncertainty of $10^{-6}$ or $10^{-7}$ would require
many more spectral features, much higher S/N in each feature, much
narrower features, or some combination of these three factors.

\begin{deluxetable}{lccrc}
  \tablecaption{Line fit results for the east image. Lines are in
    absorption except for FeII*.}
    \label{tab:linefits-east}
\tablehead{\colhead{Line} & \colhead{$\lambda_{\rm obs}$ [\AA]} & \colhead{$\sigma_\lambda$ [\AA]} & \colhead{S/N} & \colhead{width [$\sigma$,\AA]}}
\startdata
FeII & 5757.8& 0.23&7.6 &1.9 \\
FeII & 5832.5& 0.29&6.2&2.0 \\
FeII & 5852.7& 0.20&9.5&2.1 \\
FeII & 6353.2& 0.22&8.7&2.2 \\
FeII & 6386.2& 0.21&10.5 & 2.6\\
FeII*& 6451.1& 0.26&6.9 & 1.9\\
MgII& 6867.5& 0.32&9.7 & 3.3\\
MgII& 6885.4& 0.25&11.0&3.4\\
\enddata
\end{deluxetable}

In principle, the continuum could have slopes and breaks that provide
information beyond the identified lines. Our Deimos data cannot
quantify this information because they are noisy and
lack flux calibration; the apparent continuum slope in
Figure~\ref{fig-spectra} is similar to the instrument sensitivity
curve, suggesting a flat continuum. We therefore turn to a high-S/N
stacked spectrum of blue galaxies at $z{\sim}2$ \citep{Kurk2009}. At
the rest wavelength range covered by our data (2312--2813 \AA) the
continuum is indeed flat, and the lines listed in
Table~\ref{tab:linefits-east} are by far the dominant spectral features.
Therefore, template-fitting techniques are unlikely to improve the
precision in this wavelength range.

\textit{Fitting the raw data.} Information may be lost when
marginalizing 2-D slitlet spectra over the spatial dimension to obtain
1-D spectra, and/or when stacking multiple exposures into a final
product.  We investigated this by modeling the raw 2-D spectra,
including sky emission and absorption features. We found that the
wavelength precision on any given line was about the same. This
suggests that the standard PypeIt pipeline is effective at preserving
relevant information when extracting 1-D spectra and stacking them.

\subsection{Toward more informative observations}\label{subsec-anal-design}

As stated in the previous subsection, higher precision requires more
spectral lines, higher S/N lines, narrower lines, or some combination
of the three.  

Narrower lines will help, but are unlikely to provide the required
factor of ${\sim}100$. The width of the observed lines is
$\sigma\approx95$ km/s, including a subdominant amount (${\approx}50$
km/s) of instrumental broadening. For sources that are galaxies, this
is a typical velocity dispersion, so finding lines orders of magnitude
narrower is unlikely. If the source is rotating, higher spatial
resolution could resolve the rotation and reduce the apparent
dispersion by, say, a factor of three.

Therefore, most of the gain must come from additional lines and/or
higher S/N lines.  The key limiting factor in the current data is that
the S/N of absorption lines is limited by the S/N of the surrounding
continuum.  Emission lines are much brighter than the continuum and
therefore could in principle reach S/N $\approx100$.  For the lensed
galaxy observed here at $z=1.456$, [OII] $\lambda\lambda$ 3727,3729
would be redshifted to 9153 \AA, still within the reach of
ground-based silicon detectors.  We predict a very bright [OII]
doublet as follows.  First, there is a strong empirical correlation
between [OII] emission and the absorption lines we observe
\citep{Kurk2009}; recent star formation powers those ISM lines.
Second, the median [OII] equivalent width (EW) for $z{\sim}1$ galaxies
with FeII* emission is 60 \AA\ \citep{Kornei2013}.  Third, we
performed SED modeling of 7-band photometry using Prospector
\citep{JohnsonProspector2021} and found that almost all viable models
have strong [OII] emission, with a best-fit rest-frame EW of 66
\AA.\footnote{While finalizing this manuscript, strong [OII] emission
  was in fact confirmed (Hopp et al in preparation).}  As befits a
star-forming galaxy, these models also predict strong Balmer lines
further into the infrared. Instruments that cover the 1--1.6 $\mu$m
range would capture H$\delta$ through H$\alpha$.

\subsection{Testing consistency across lines}

\begin{figure}
\centerline{\includegraphics[width=\columnwidth]{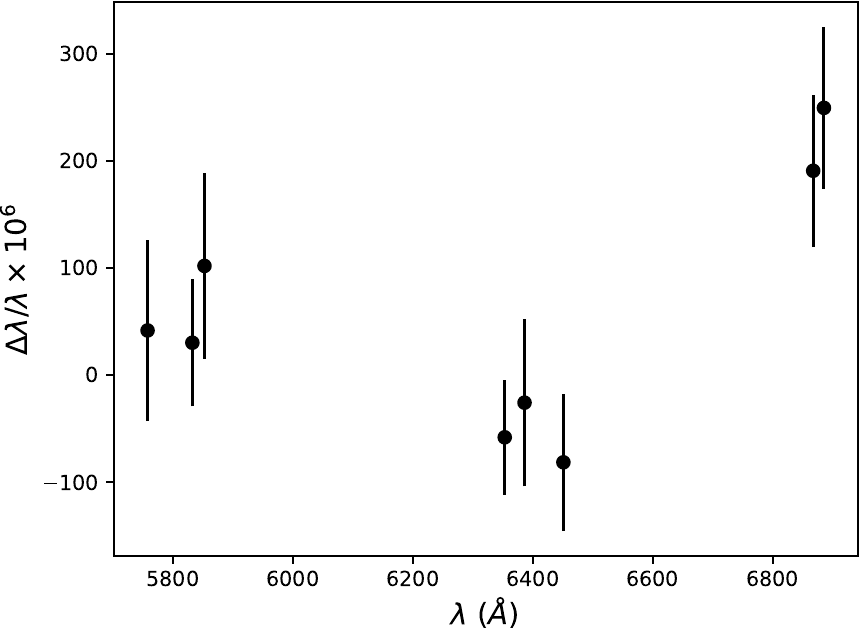}}
\caption{$\dlol$ as measured between the east and central images for
  each of the eight spectral lines. The wavelength grouping indicates
  that wavelength calibration variations using the standard pipeline
  are of the order of hundreds of parts per million.}
\label{fig-lineratios}
\end{figure}

\S\ref{subsec-anal-stat} cited only the statistical uncertainties from
the line fitting. Here we ask whether the various spectral lines are
consistent. Returning to the 1-D spectra produced by PypeIt, we fit
for $\dlol$ between east and central images.  This statistic is
constructed to be immune to intrinsic offsets, say between Mg and
FeII, because each line is compared across the two images
independently of the other lines.  Figure~\ref{fig-lineratios} plots
the results as a function of observed wavelength. The lines fall into
clear groups based on wavelength, suggesting that wavelength
calibration variations using the standard pipeline are of the order of
hundreds of parts per million.

The observing setup and pipeline were not designed for the level of
precision required to measure the moving lens effect. The two lensed
images were observed through different slitmasks and PypeIt uses arc
lamp exposures, unique to each slitmask, for wavelength
calibration. PypeIt also has a flexure model which could introduce
exposure- and slit-dependent differences after the arc fitting stage.
This approach is not well suited to the moving lens use case, which is more
sensitive to calibration errors that are different between the lensed
images than to absolute calibration errors.  The differential error
would likely be substantially smaller if the two images had been
observed simultaneously through one long slit.

That said, controlling wavelength calibration errors at the level
required to detect the moving lens effect is still a serious
challenge. Spectrographs designed for precise stellar radial
velocities, for example in exoplanet searches, reach even better
precision but are limited to much brighter targets.  However, the
moving lens use case does not require that observations \textit{widely
  separated in time} must be put on the same footing; rather it
requires that \textit{two different targets, possibly observed
  simultaneously} share the same footing.  Sky emission lines, which
have the virtue of being observed simultaneously with the galaxy
images, could provide the appropriate wavelength
references. 

\begin{figure}
\centerline{\includegraphics[width=\columnwidth]{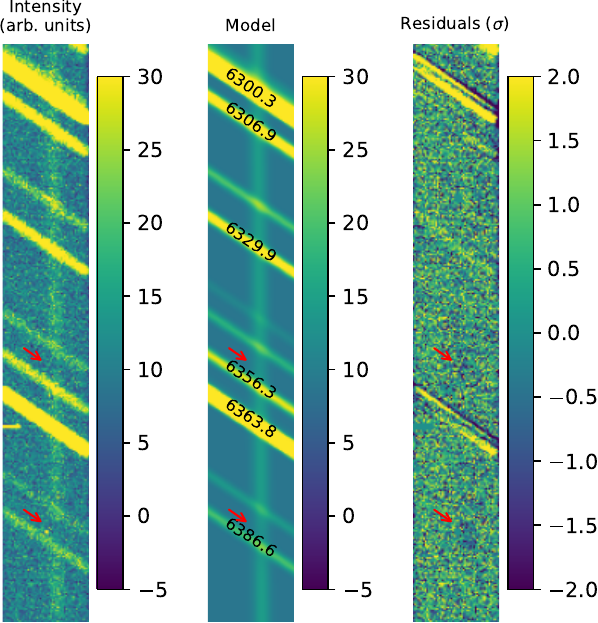}}
\caption{A small part of one exposure of the two-dimensional spectrum:
  data (left), model (middle), and residuals (right). The FeII
  $\lambda\lambda 2586.65,2600.17$ (rest frame) galaxy absorption
  lines are marked with arrows, and the sky emission lines are marked
  with their wavelengths as tabulated in the UVES atlas of optical sky
  emission lines \citep{Hanuschik2003}. Cosmic rays were masked while
  fitting, hence do not appear in the model or residuals.}
\label{fig-2dfit}
\end{figure}

We next ask whether calibration with sky emission lines would offer
sufficient statistical precision.  Because sky lines are automatically
removed in the PypeIt processing, we return to the raw 2-D spectra for
the east image. Because the galaxy lines occur in widely separated
groups, we choose small wavelength regions around each group, over
which a simple linear wavelength solution should apply.
Figure~\ref{fig-2dfit} shows the observed wavelength range 6297–-6407
\AA, spanning the FeII $\lambda\lambda 2586.65,2600.17$ absorption
lines. Using the UVES atlas of optical sky emission lines
\citep{Hanuschik2003} to assign wavelengths to observed sky lines, we
simultaneously fit for the wavelength solution and the shift of the
galaxy absorption lines relative to the sky lines. This method yields
uncertainties in $\dlol$ similar to those found above with fixed
wavelength solutions. Hence, we conclude that deriving the wavelength
solution from sky emission lines will contribute subdominantly to the
statistical error budget (systematics are briefly considered in
\S\ref{ssec-confound}). This statement is not generalizable because
the number and intensity of sky lines varies greatly with wavelength.
Nevertheless, this particular lensed galaxy is a promising candidate
because its expected emission lines are in a wavelength region rich
with sky lines.

\section{Potential impact on cluster merger modeling}\label{sec-mergermodeling}

% DW: note somewhere here that we're not considering substructure

\subsection{Basic approach}

\begin{figure}
\centerline{\includegraphics[width=\columnwidth]{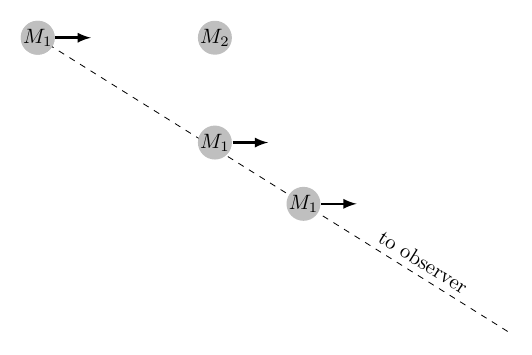}} 
\caption{Impact of not knowing the relative LOS distance of two
  subclusters, even when LOS and POS motions of $M_1$ can be measured
  and the full velocity vector reconstructed (arrows). Depending
  on where it is along the LOS, $M_1$ could be approaching $M_2$
  (upper left position), receding from $M_2$ (lower right position),
  or neither (middle position; with velocity perpendicular to the
  separation vector, this could be pericenter or apocenter).}
\label{fig-LOS}
\end{figure}

The interpretation of subcluster velocities is complicated by our
inability to determine which subcluster is closer to the observer, as
illustrated in Figure~\ref{fig-LOS}.  Suppose that LOS and POS motions
of $M_1$ can be measured (using $M_2$ as a reference for both position
and velocity), and the full velocity vector reconstructed (black
arrows). Depending on where it is along the LOS, $M_1$ could be
approaching $M_2$ (upper left position), receding from $M_2$ (lower
right position), or neither. In the middle position shown, the
velocity is perpendicular to the separation vector, so this could be
pericenter (if the speed is high and the separation small) or
apocenter (if the speed is low and the separation large).  Weighing
these scenarios requires prior information on the distribution of,
e.g., pericenter distances and speeds in a $\Lambda$CDM universe.

We therefore build on the method of \citet{Wittman19analogs}, who
identified major mergers in the halo catalog of the BigMDPL
\citep{BigMDPL2016} cosmological simulation. The large box size (2.5
Gpc/h)$^3$ of this simulation enables the identification of a variety
of analog mergers (halo pairs after first pericenter) capable of
fitting any particular observed scenario.  The basic observables in
the original implementation are projected separation $d_{\rm proj}$
and the line-of-sight (LOS) velocity difference $\Delta v_{\rm los}$
between the two clusters, and the mass of each subcluster.

Mock observers are placed at random viewing angles and a likelihood is
computed for the analog system and viewing angle, which we
collectively refer to as the model. The viewing angle $\theta$ is
defined as the angle between the LOS and the current subcluster
separation vector. Hence, $\theta=90^\circ$ indicates a merger with a
separation vector in the plane of the sky while $\theta=0^\circ$ is a
merger along the LOS. (Models at fixed $\theta$ are not all identical
due to variation of the azimuthal viewing angle, which is not shown in
this paper.) The angle $\varphi$ between the current separation and
velocity vectors is also tabulated as a proxy for merger phase
($\varphi >90^\circ$ for inbound systems). Following
\citet{Wittman19analogs}, we limit the analogs to those between first
and second pericenter, because those are the most easily identifiable
mergers based on X-ray properties.

% ; the latter scenario is likely to be selected against due to the
% lack of clearly identifiable subclusters in sky surveys.

\subsection{Sky-plane velocities}

We modify this code to compute the plane-of-sky motions as well.  We
noted in \S\ref{sec-intro} that strongly lensed images are more likely
to be split perpendicular to the separation vector because a merging
system resembles a bar of mass. If these images are common, it is
worth investigating the speeds to be expected in that direction, in
addition to the motion along the separation vector, which is
presumably faster but less likely to have images well placed to probe
it.  To reflect these possible lensing geometries, we separate the
plane-of-sky motion into components parallel and perpendicular to the
projected separation vector, which we denote  $v_{\rm sky,\parallel}$ and
$v_{\rm sky,\perp}$ respectively.

There are some important distinctions between these transverse
velocities and the $\Delta v_{\rm los}$ that is observed using the
Doppler effect on member galaxies:
\begin{itemize}
\item The sign on $\Delta v_{\rm los}$ is arbitrary because one
  subcluster is used as a velocity reference for the other, and the
  LOS separation is unknown.  A positive (negative)
  $v_{\rm sky,\parallel}$, in contrast, meaningfully indicates that
  the subclusters are approaching (receding from) each other in the
  sky plane. Although this is not a statement about the orbit in 3-D,
  it helps rule out many models as illustrated below.  The sign of
  $v_{\rm sky,\perp}$ remains arbitrary.
\item The sky-plane velocity is not a differential measurement.  There
  are \textit{two} moving lenses and each could in principle be
  measured independently.  The lower-mass subcluster is likely to
  provide larger $\dlol$ because its velocity (relative to the system
  center of mass) scales as $M^{-1}$ while in a simple point-mass
  approximation the Einstein angle (a rough proxy for the deflection
  angle) scales as $M^{1/2}$, hence $\dlol\propto M^{-1/2}$.  The
  measurements available will be dictated by chance image placement
  (and will probe the sky-plane velocity at random position angles
  rather than splitting neatly into $v_{\rm sky,\parallel}$ and
  $v_{\rm sky,\perp}$).  To illustrate the potential impact of
  sky-plane velocity measurements without overly focusing on a
  particular scenario, we generate the distribution of
  $v_{\rm sky,\parallel}$ and $v_{\rm sky,\perp}$ for the lower-mass
  subcluster in each analog, conditioned on mock measurements of the
  other observables described below.  We assume the source and
  observer are stationary in the BigMDPL comoving coordinate system,
  so we use the subcluster velocity directly as listed in the BigMDPL
  database.
\end{itemize}

\begin{figure*}
\centerline{\includegraphics[width=0.9\textwidth]{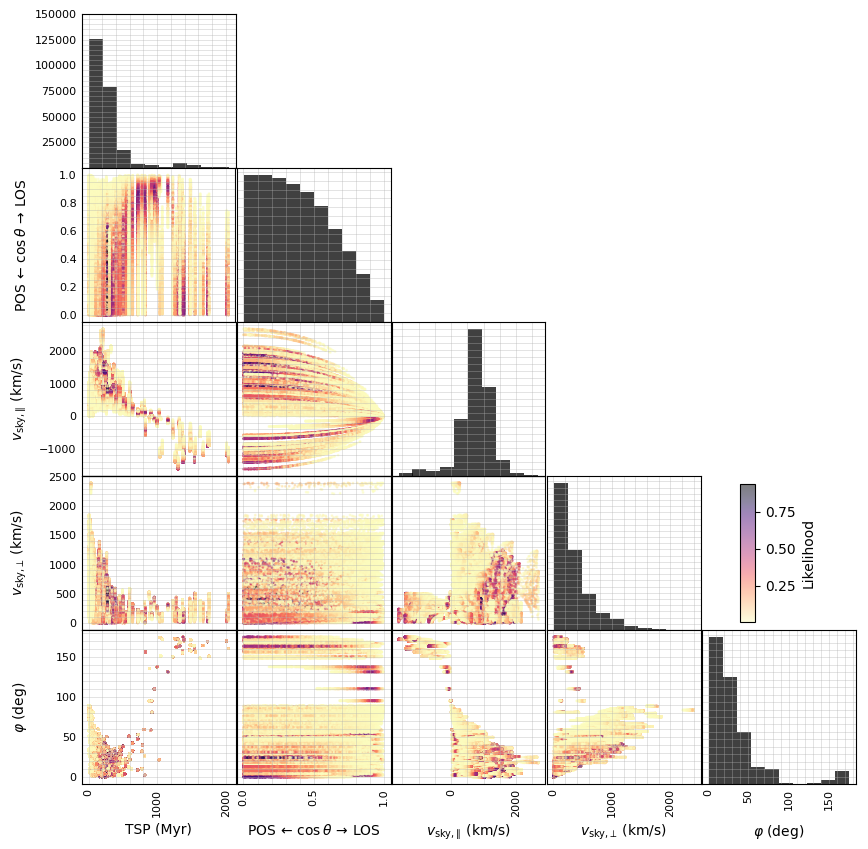}}
\caption{Corner plot showing relationships between time since
  pericenter (TSP); angle $\theta$ between the LOS and the current 3-D
  separation vector; sky-plane velocity components (for the lower-mass
  subcluster) parallel and perpendicular to the projected separation
  vector between the two merging subclusters; and angle $\varphi$,
  between the 3-D separation and velocity vectors, which is
  $>90^\circ$ for inbound systems. Each point is a model (an analog
  system combined with a value of $\theta$) color-coded by likelihood
  given the generic observations listed in the text. The POS and LOS
  labels on the $\cos\theta$ panels refer to the current 3-D
  separation vector being in the POS or along the LOS.}
\label{fig-corner}
\end{figure*}

For illustration purposes, we assume a generic merger with observed
values of $d_{\rm proj}=500\pm50$ kpc, $\Delta v_{\rm los}=100\pm300$
km/s, and subcluster masses of $5\pm2\times10^{14}$ M$_\odot$.  For each
model, we compute the likelihood and then plot that model as a point
in Figure~\ref{fig-corner} color-coded by
likelihood. With 804 analog systems, each viewed from 500 random
angles, there are 402,000 points, so each point is plotted at low
opacity. Figure~\ref{fig-corner} includes two model attributes that
depend solely on the simulated analog (time since pericenter or TSP,
and $\varphi$) and one model parameter that
depends solely on the observer's LOS ($\cos\theta$).  It also includes the two potential
observables $v_{\rm sky,\parallel}$ and $v_{\rm sky,\perp}$, which
depend on both the analog and the LOS. By
disregarding color, the reader can see the relationships among all
BigMDPL merging clusters at this snapshot (representing $z=0.2947$),
and by focusing on the darker colors the reader can see the subset
that best match the mock observables listed at the start of this paragraph.

% To trim a png: magick corner.png -trim +repage -bordercolor white -border 10 output.png

% INTERPRET
The notable patterns in Figure~\ref{fig-corner} are:
\begin{itemize}
\item \textit{Monotonic relationship between $v_{\rm sky,\parallel}$
    and TSP.} Figure~\ref{fig-LOS} showed that, when stripped of
  cosmological context, even perfect knowledge of all three components
  of the velocity vector is not enough to disambiguate the merger
  stage. The $v_{\rm sky,\parallel}$-TSP panel of
  Figure~\ref{fig-corner} shows that in the context of $\Lambda$CDM,
  it \textit{can be} enough. In that panel, $v_{\rm sky,\parallel}$
  changes gradually from +$2000$ km/s at pericenter to -$1500$ km/s
  for the oldest systems, which are close to second pericenter as
  confirmed by the $\varphi$ panels. Focusing on the high-likelihood
  systems, the typical $v_{\rm sky,\parallel}$ drops by
  about 500 km/s per 200 Myr past pericenter, so even a noisy observation of
  $v_{\rm sky,\parallel}$ could be useful in disambiguating between
  outbound, near-apocenter (${\approx}900$ Myr), and returning scenarios.
  At the same time, there is a scatter of hundreds of Myr in TSP at
  fixed $v_{\rm sky,\parallel}$, so even an exact measurement of
  $v_{\rm sky,\parallel}$ would not fix TSP precisely.
  
\item \textit{$v_{\rm sky,\perp}$ not always negligible.}  For this
  set of mock observables, $v_{\rm sky,\perp}$ reaches ${\approx}1000$
  km/s, and a handful of models (which are poor matches to these mock
  observables) even reach 2000 km/s.  This is perhaps surprisingly
  high and may motivate more spectroscopy of the common scenario in
  which strongly lensed images are split perpendicular to the major
  axis of the mass distribution of the merging system.  These speeds
  happen near pericenter, making for a good match to systems selected
  based on X-ray morphology, which is distinctive soon after
  pericenter.  That said, only a small minority of systems, even of
  those with low TSP, have $v_{\rm sky,\perp}$ approaching 1000 km/s.
  To rule out half the models shown, one would have to place an upper
  limit of 200 km/s on $v_{\rm sky,\perp}$, which would be extremely
  challenging observationally. The interpretation would also be
  complicated because at this level the motions of source and
  observers are no longer negligible.

\item \textit{U-shaped relationships involving $\cos\theta$.} These
  are related to the mock observation $d_{\rm proj}=500\pm50$
  kpc. This is much smaller than a typical apocenter distance, so near
  apocenter models must
  foreshorten the separation with $\cos\theta$ near unity. Likewise,
  pericenter distances near 500 kpc are common, so near first or
  second pericenter, models must have  $\cos\theta$ near zero.  This
  creates a U-shaped relationship between $\cos\theta$ and TSP,
  which in turn creates a U-shaped relationship between $\cos\theta$ and 
 $v_{\rm sky,\parallel}$ due to the monotonic relationship between
 TSP and $v_{\rm sky,\parallel}$.
  \item \textit{Only mild correlations between $v_{\rm sky,\perp}$ and
    the other parameters.} The upper envelope of the
  $v_{\rm sky,\perp}$ distribution declines with TSP; if one were to
  observe $v_{\rm sky,\perp}\approx1000$ km/s it would very much
  narrow the range of possible TSP for the system. However, measuring
  a typical $v_{\rm sky,\perp}$ of 0--500 km/s would provide no
  information on TSP. Because of the linkage between TSP and $\theta$,
  a similar statement applies to constraining $\theta$ based on $v_{\rm sky,\perp}$.
  There is no clear correlation between $v_{\rm sky,\perp}$ and $v_{\rm
    sky,\parallel}$ except that a small magnitude of $v_{\rm
    sky,\parallel}$ implies a small magnitude of $v_{\rm
    sky,\perp}$.
\end{itemize}

The implication for RM J001938.7+033557.3 is that the maximum
predicted $\dlol$ between south and central images is $4\times10^{-7}$
(given the small image separation and maximum $v_{\rm sky,\parallel}$
of 2000 km/s), while the east and west images could reach twice that,
given their much larger separation and that $v_{\rm sky,\perp}$ is at
or above 1000 km/s for some reasonably likely models.

\section{Summary and discussion}\label{sec-discuss}

\subsection{Summary}
We have discovered a galaxy quadruply imaged by the southern
subcluster of the merging galaxy cluster RM J001938.7+033557.3. We
presented spectra of two of the images and confirmed their consistency
at $z=1.4561$. We argue that this system is a good candidate for
eventually detecting the moving lens effect: the images are relatively
bright ($r=22$) for lensed galaxies; they appear in a useful geometry
on the sky (with two of the images split largely along the presumed
direction of motion if the merger is head-on); and merging clusters
maximize the amplitude of the effect because they are massive and move
at high speeds.

Although the spectra were not designed to measure the moving lens
effect, we performed an initial analysis to highlight the issues
involved in such an analysis, and to motivate future observations more
likely to support a useful upper limit and eventually a detection. We
argue that optical spectroscopy must focus on sources with very strong
emission lines: with a typical galaxy linewidth of $\sim100$ km/s,
S/N$>100$ will be needed to reach sub-km/s precision. We also
presented evidence that the source considered here will have strong
lines in the infrared.  Controlling systematic error, particularly
wavelength calibration, will be a challenge for future work.

We analyzed simulated merging clusters to reveal the distribution of
sky-plane motions in a $\Lambda$CDM cosmology, and their correlations
with merger model parameters. We found a usefully tight and monotonic
relationship between $v_{\rm sky,\parallel}$ (the sky-plane motion
measured parallel to the sky-plane separation vector) and two probes
of merger phase: TSP and $\varphi$, the angle between the 3-D
separation and velocity vectors. This suggests that even coarse
measurements of $v_{\rm sky,\parallel}$ could usefully narrow the
range of merger models for a given observed system.  Because most
images lensed by merging clusters are split perpendicular to the
projected subcluster separation vector, we also tabulated
$v_{\rm sky,\perp}$. It is typically a few hundred km/s, making its
detection even more challenging, but upper limits of 1000 km/s could
rule out some of the more extreme models.

\subsection{Possible confounding factors}\label{ssec-confound}

It is worth asking whether other physical effects could produce a
$\dlol$ effect of order $10^{-6}$ and mimic the moving lens
effect. These can be divided into observational difficulties and
additional physical effects.

\textit{Observational difficulties.} Extended sources will present not
a single velocity but a velocity field, with spatial variations of
order $\sim100$ km/s for rotation-supported galaxies.  An accurate
lens model will be required to compare the relevant part of each
velocity field across images.  The magnification will vary across each
image, so the integrated light of different images may be dominated by
different parts of the galaxy, with different velocities. This effect
was identified and termed the Differential Magnification Effect (DME)
by \citet{BanikZhao2015}, who suggested minimizing it by selecting
sources with more uniform velocity fields.  We suggest that bright and
well-placed images are rare enough to merit observing the velocity
fields with integral field units (IFUs) at high angular resolution.
The DME will be mitigated if the resolution elements are fine enough
to capture roughly uniform-velocity patches of the source and the lens
model is accurate enough to match those patches across images. If the
patches are larger, an accurate lens model would still enable an
estimate of the DME to be incorporated into the inference.

% more about wvlngth calib
Wavelength calibration may be the largest observational
hurdle. Although absolute calibration is not needed, the variation in
calibration from one image to the other must be controlled to a
precision of $10^{-7}$ or better, and most faint-object spectrographs
are not designed to do this.  For ground-based facilities, sky lines
imprinted on the galaxy spectrum can provide simultaneous calibration,
but sky emission lines can still trigger a small systematic error when
a target is spatially offset from the center of the slit. Sky
absorption lines imprinted on the target spectrum are immune to this
effect and are used for velocities of stars precise to the sub-km/s
level \citep{Geha2026}. Space-based facilities may offer better
angular resolution to mitigate the DME, but have no comparable
independent check on the wavelength calibration.
  
\textit{Physical effects.} Because of time delays, two lensed images
show the same source galaxy but at different times. Time delays for
cluster-scale lenses are of order years, up to a decade. If the source
galaxy accelerates during that decade, the temporal velocity change
could mimic the moving lens effect.  A worst-case scenario could be
that the source galaxy is orbiting or falling into a much more massive
galaxy. For comparison, given the Milky Way's mass and distance from
the Large Magellanic Cloud (LMC), the LMC acceleration is about
$2.5\times10^{-11}$ m s$^{-2}$, yielding a velocity difference of
about 1 cm s$^{-1}$ after one decade. This would yield
$\dlol\approx 3\times 10^{-11}$, far smaller than the moving lens
effect considered here. For $\dlol$ measurements based on emission
lines, one could argue that only the (possibly small) emission line
region needs to accelerate to create a confounding situation.  As a
worst-case scenario, let the emitting region be comparable to the Sun
in its Milky Way orbit: the acceleration is an order of magnitude
larger than for the LMC, but even with a decade of time delay the
velocity difference between images is still much smaller than that
caused by the moving lens.

% Possibly useful: Table 3 in https://arxiv.org/abs/1911.01467
% quantify various Sun/MW accelerations.

Intrinsic variability is another possibility.  Outbursts from emitting
regions could potentially change the observed velocity over time, but
only very rapid changes (1 km/s over a decade) would cause an issue
given the timescale of cluster lens delays.  Furthermore, these
changes would have to affect not just a small region, but most of
a lensed image. A potential source
of false positives is AGN activity. Sources should be checked for AGN
signatures---but AGN are not necessarily disqualified as sources,
as discussed in the next subsection.

% Not needed: Previous studies of galaxy variability have focused on
% line \textit{strengths} and found that such variation is rare
% \citep{Lin_2022}. Even these would have no effect on the $\dlol$
% measurement.

\citet{BanikZhao2015} calculated the increased cosmological redshift
imposed on one image due to the time delay. They conclude that the
moving lens effect is 1000 times larger than this differential
expansion effect.

Source and observer motions are assumed to be subdominant here, but
this may not always be the case: even a merging cluster may have
little sky-plane motion at certain phases and viewing angles.
\citet{Wucknitz2004} found that the contribution of source velocity is
scaled by $\frac{1+z_d}{1+z_s}\frac{D_{\mathrm d}}{D_{\mathrm s}}$
relative to the lens velocity contribution; each ratio in that
expression is strictly less than one, and the product is 0.25 for RM
J001938.7+033557.3. Hence a source speed of 400 km/s could change the
inferred lens velocity by up to 100 km/s, depending on the alignment
of the motions and the image splitting. The same authors found that
observer velocity is scaled by $\frac{D_{\mathrm ds}}{D_{\mathrm s}}$
relative to the lens velocity contribution; this fraction is also
strictly less than one and is 0.75 for RM J001938.7+033557.3.  Hence
the Sun's roughly 370 km/s velocity relative to the CMB
\citep{Planck2020} could bias the inference by up to a few hundred
km/s if not corrected for.

\subsection{Potential future directions}

\textit{Other types of sources.} QSOs have very strong emission lines,
and some of them are considered narrow. However, ``narrow'' in this
context means 300-1000 km/s, which implies S/N requirements 3--10
times greater than for a typical galaxy.  QSOs are also time variable,
which combined with different time delays for the different images,
could make spectra differ even without the moving lens effect. On the
other hand, while QSO lines vary in strength, it is not clear how much
they vary in velocity.  Future work is needed to explore these issues,
but only about ten QSOs are known to be strongly lensed by clusters
\citep{Bazzanini2025}, of which merging clusters are a subset.

\textit{Other types of lines.} The velocity dispersion of molecular
gas is 2.5--3 times lower than that of atomic gas in galaxies at this
redshift \citep{Girard2021} so the suggestion by \citet{Molnar2013} to
scan molecular lines using ALMA could lower the S/N requirement by
the same factor. ALMA is also capable of excellent spatial resolution
for controlling DME as described above. Perhaps most importantly,
the ALMA frequency scale is electronically injected and is accurate to
mm/s precision, thus eliminating the wavelength calibration issue.

% \citet{Girard2021} says σ0,mol ∼ 8–27 km s−1 and σ0,ion ∼ 25–71 km
% s−1 (sample includes "z ∼ 1–2 Disks and Local Analogs". So ratio is
% 2.6-3.1. (Doesn't make sense to use their 2.45 after correcting for
% greater thermal broadening in ionized gas.) 

\textit{Astrometric signal.} Even to the extent that a 3,000 km/s lens
could be seen directly (for example via its association with a BCG),
at 1 Gpc from us its proper motion would be less than one microarcsec
per year. The placement of strongly lensed images is more sensitive to
the lens-source alignment. A simple estimate of the increased
sensitivity factor using axisymmetric lenses yields
$\frac{\theta}{\beta}$, which could provide a large factor for rare
favorably placed sources. Each cluster is a sufficiently complicated
lens that the sensitivity of image placement (and magnification) to
its motion should be examined individually using its best-fit lens
model.

% Consider an axisymmetric lens moving east with a source $\beta$
% arcsec directly north, appearing at $\theta$ arcsec no

% the simple case of a point lens, where the two images appear at sky
% positions
% $\theta_{\pm}=\frac{1}{2}(\beta\pm\sqrt{\beta^2+4\theta_E^2})$
% \citep{NarayanBartelmann} where $\theta_E$ is the Einstein angle and
% $\theta$ and $\beta$ are one-dimensional versions of the angles
% defined in \S\ref{sec-intro}, reflecting the axisymmetry of the situation.

  \textit{Application to microlensing.} In microlensing, $\dlol$ will
  be much smaller, but the sources are stars that are substantially
  brighter than the galaxy source considered here, possibly accessible
  with planet-hunting spectrographs capable of reaching 0.1 m/s
  accuracy. For a source in the Galactic bulge and a solar-mass lens
  midway between observer and source, deflection angles are roughly a
  milliarcsec. At $v=200$ km/s this produces $\dlol=5\times10^{-12}$
  or a velocity equivalent of order mm/s, well below current
  spectroscopic capabilities. This scales as $M^{1/2}$, so even 100
  $M_\odot$ lenses are insufficient to produce a detectable signal.

  A second consideration is that the two images are unresolved.  This
  will produce a composite spectrum with some photons blueshifted and
  some redshifted, but the signal will not completely cancel because
  one image is magnified more than the other.  Using the simple ansatz
  that the apparent line shift in the composite spectrum is the sum of
  the individual shifts weighted by their magnifications, we find that
  the maximum composite $\dlol$ is about 0.7 times that expected from
  the brighter image alone. Hence the main factors impeding the
  microlensing application are the small deflections and velocities,
  rather than the image blending.

  % \textit{Combination with other probes.} ??  CH says: Maybe some
  % mention of kSZ and a combined uncertainty on a 3d velocity But: we
  % already have LOS msmnts of the GALAXIES (thought to be better than
  % the gas) and who cares about v_3d by itself, we care about orbit
  % modeling.

  \textit{Robustness tests.} If future ground-based observations seem
  to reach the required $\dlol$ precision of $10^{-7}$-$10^{-6}$,
  observatory motion could be quite useful as a robustness
  test. Earth's rotation over the course of a night would provide a
  shift of a few parts per million relative to the sky lines. The same
  shift would be imprinted on both lensed images, so the $\dlol$
  between simultaneously observed images should remain the same
  throughout.

\acknowledgments This work was supported by NSF grant number 2308383.
We thank Faik Bouhrik for bringing this cluster to our attention, and
Zhuoran Gao for reducing the raw Deimos spectra to 1-D
wavelength-calibrated spectra using \texttt{PypeIt}. We thank Wonki
Lee, James Jee, Tucker Jones, and Srinivasan Raghunathan for useful
discussions.  Some of the data presented herein were obtained at Keck
Observatory, which is a private 501(c)3 non-profit organization
operated as a scientific partnership among the California Institute of
Technology, the University of California, and the National Aeronautics
and Space Administration. The Observatory was made possible by the
generous financial support of the W. M. Keck Foundation.  The authors
wish to recognize and acknowledge the very significant cultural role
and reverence that the summit of Maunakea has always had within the
Native Hawaiian community. We are most fortunate to have the
opportunity to conduct observations from this mountain.

\facilities{Keck:II (DEIMOS)} 

\software{This research made use of community-developed or
  community-maintained software packages, including Matplotlib
  \citep{matplotlib2007}, NumPy \citep{numpy2020}, and SciPy
  \citep{scipy2020}.  Deimos spectra were calibrated and reduced with
  PypeIt \citep{pypeit:joss_pub,pypeit:zenodo}.  This research has
  also made use of NASA’s Astrophysics Data System.}

The {\it HST} images used in Figures 1 and 2 are available at MAST:
\dataset[doi: 10.17909/dwna-jq08]{\doi{10.17909/dwna-jq08}}.

\bibliography{ms}

\end{document}